%% file: main.tex
\documentclass[sigconf]{acmart}
\usepackage{multirow}
\usepackage{tabularx}   
\usepackage{booktabs}   
\usepackage{multirow}   
\usepackage{subcaption}
\usepackage{enumitem}
\usepackage{tcolorbox}
\AtBeginDocument{%
  }

\setcopyright{acmlicensed}
\copyrightyear{2018}
\acmYear{2018}
\acmDOI{XXXXXXX.XXXXXXX}
\acmConference[Conference acronym 'XX]{Make sure to enter the correct
  conference title from your rights confirmation email}{June 03--05,
  2018}{Woodstock, NY}
\acmISBN{978-1-4503-XXXX-X/2018/06}

\acmSubmissionID{209}

\begin{document}

\title{MUSE: A Simple Yet Effective Multimodal Search-Based Framework for Lifelong User Interest Modeling}

\author{Bin Wu}
\authornote{Bin Wu and Feifan Yang contributed equally to this research. Work done by Bin Wu during internship at Alibaba Group.}
\email{wubin2021@whu.edu.cn}
\affiliation{%
  \institution{Wuhan University}
  \city{Wuhan}
  \state{Hubei}
  \country{China}
}

\author{Feifan Yang}
\authornotemark[1]
\email{yangfeifan.yff@alibaba-inc.com}
\affiliation{%
  \institution{Alibaba Group}
  \city{Beijing}
  \country{China}
}

\author{Zhangming Chan}
\email{zhangming.czm@alibaba-inc.com}
\affiliation{%
  \institution{Alibaba Group}
  \city{Beijing}
  \country{China}
}

\author{Yu-Ran Gu}
\email{guyuran.gyr@alibaba-inc.com}
\affiliation{%
  \institution{Alibaba Group}
  \city{Beijing}
  \country{China}
}

\author{Jiawei Feng}
\email{jwf3ng@gmail.com}
\affiliation{%
  \institution{Alibaba Group}
  \city{Beijing}
  \country{China}
}

\author{Chao Yi}
\email{yunan.yc@alibaba-inc.com}
\affiliation{%
  \institution{Alibaba Group}
  \city{Beijing}
  \country{China}
}

\author{Xiang-Rong Sheng}
\email{xiangrong.sxr@alibaba-inc.com}
\affiliation{%
  \institution{Alibaba Group}
  \city{Beijing}
  \country{China}
}

\author{Han	Zhu}
\email{zhuhan.zh@alibaba-inc.com	}
\affiliation{%
  \institution{Alibaba Group}
  \city{Beijing}
  \country{China}
}

\author{Jian Xu}
\email{xiyu.xj@alibaba-inc.com}
\affiliation{%
  \institution{Alibaba Group}
  \city{Beijing}
  \country{China}
}

\author{Mang Ye}
\authornote{Mang Ye and Bo Zheng are the corresponding authors.}
\email{yemang@whu.edu.cn}
\affiliation{%
  \institution{Wuhan University}
  \city{Wuhan}
  \state{Hubei}
  \country{China}
}

\author{Bo Zheng}
\email{bozheng@alibaba-inc.com}
\authornotemark[2]
\affiliation{%
  \institution{Alibaba Group}
  \city{Beijing}
  \country{China}
}

\renewcommand{\shortauthors}{Wu et al.}

\begin{abstract}
Lifelong user interest modeling is crucial for industrial recommender systems, yet existing approaches rely predominantly on ID-based features, suffering from poor generalization on long-tail items and limited semantic expressiveness. While recent work explores multimodal representations for behavior retrieval in the General Search Unit (GSU), they often neglect multimodal integration in the fine-grained modeling stage---the Exact Search Unit (ESU). In this work, we present a systematic analysis of how to effectively leverage multimodal signals across both stages of the two-stage lifelong modeling framework. Our key insight is that simplicity suffices in the GSU: lightweight cosine similarity with high-quality multimodal embeddings outperforms complex retrieval mechanisms. In contrast, the ESU demands richer multimodal sequence modeling and effective ID–multimodal fusion to unlock its full potential. Guided by these principles, we propose MUSE, a simple yet effective multimodal search-based framework. MUSE has been deployed in Taobao display advertising system, enabling 100K-length user behavior sequence modeling and delivering significant gains in top-line metrics with negligible online latency overhead. To foster community research, we share industrial deployment practices and open-source the first large-scale dataset featuring ultra-long behavior sequences paired with high-quality multimodal embeddings. Our code and data is available at \url{https://taobao-mm.github.io}.
\end{abstract}




\maketitle

\input{sections/1.Introduction}
\input{sections/2.Preliminaries}
\input{sections/3.Insights}
\input{sections/4.Method}

\input{sections/5.Deployment}
\input{sections/6.Dataset}
\input{sections/7.Experiments}
\input{sections/8.Related}
\input{sections/9.Conclusion}
\input{sections/10.Acknowledgements}



\begin{figure*}[ht!]
    \centering
    \includegraphics[width=0.7\linewidth]{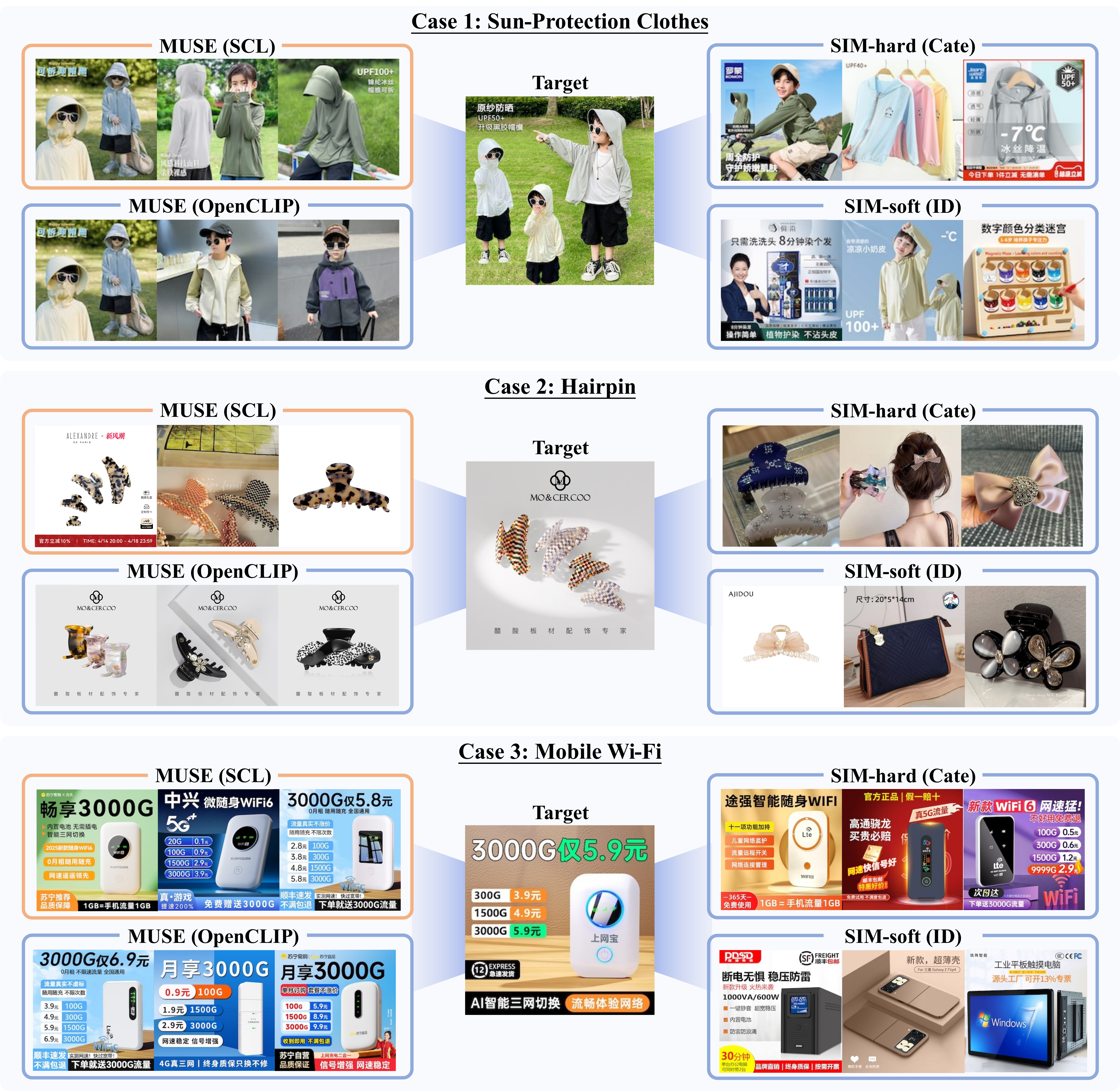}
    \caption{Cases of how searching results can be different among different GSU types.}
    \label{fig:gsu_case}
\end{figure*}

\bibliographystyle{ACM-Reference-Format}
\balance
\bibliography{main}

\appendix
\input{sections/11.Appendix}

\end{document}

%% file: sections/1.Introduction.tex
\begin{figure*}[!t]
\centering 
\includegraphics[width=\textwidth]{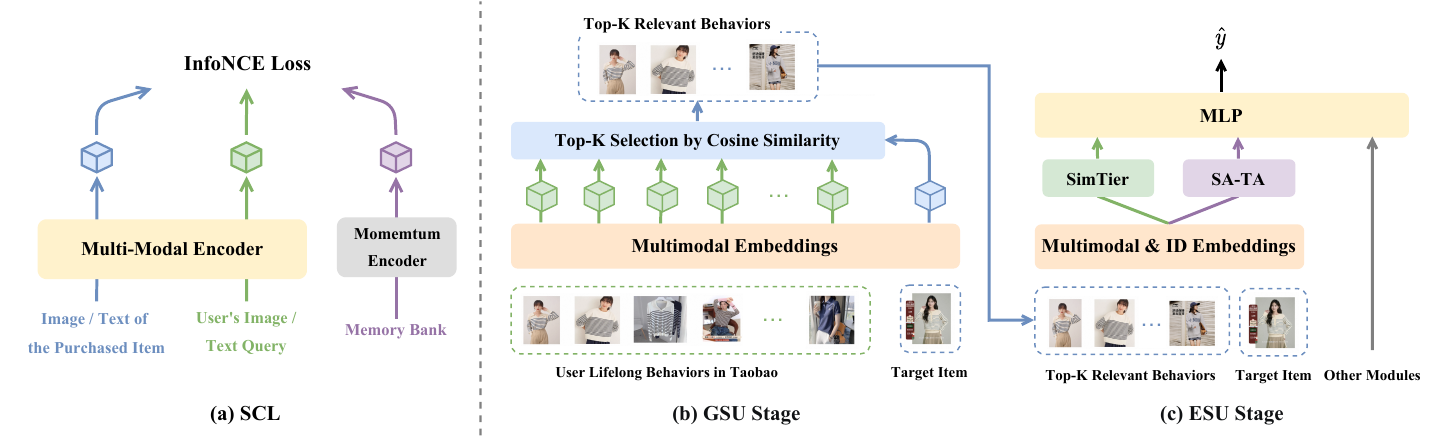} 
\caption{Overview of MUSE. (a) Multimodal item embeddings are pre-trained via Semantic-aware Contrastive Learning (SCL). In the recommendation phase,  (b) the GSU stage efficiently retrieves the top-$K$ behaviors most relevant to the target item from the user’s lifelong history using lightweight multimodal cosine similarity, drastically reducing the sequence length for downstream processing. (c) The ESU stage models fine-grained user interests through two components: the SimTier module compresses multimodal similarity sequences into histograms, while the Semantic-Aware Target Attention (SA-TA) module enriches ID-based attention with semantic guidance to produce the final lifelong user interest representation.} 
\label{fig:framework} 
\end{figure*} 
\section{Introduction}
\label{sec:introduction}

Click-Through Rate (CTR) prediction~\cite{DIN,DIEN,DSIN,BST,BianWRPZXSZCMLX2022CAN} plays a critical role in industrial recommender systems. 
As user behavior logs accumulate over time---often exceeding $10^5$ behaviors per user---they contain rich, evolving interest patterns. To leverage such lifelong sequences, modern CTR prediction models adopt a two-stage architecture~\cite{SIM,TWIN}:  
(1) a \textbf{General Search Unit (GSU)} retrieves a short, relevant subsequence from the full historical behaviors;  
(2) an \textbf{Exact Search Unit (ESU)} performs fine-grained user interest modeling on this subsequence for final prediction.

However, existing approaches rely almost exclusively on ID-based features in both stages~\cite{SIM,TWIN}. This leads to two critical limitations:  
(1) ID embeddings of long-tail or outdated items are poorly learned, degrading GSU’s retrieval quality;  
(2) the ESU lacks semantic expressiveness, as it cannot generalize beyond co-occurrence signals.  
While recent work (e.g., MISS~\cite{MISS}) introduces multimodal representations into the GSU, it still models only ID features in the ESU, leaving the second limitation unaddressed.

In this work, we systematically investigate \textbf{how to fully leverage multimodal information to enhance lifelong user interest modeling}, with a focus on integrating multimodal representations into \textit{both} the GSU and ESU. Through extensive experiments on industrial-scale data, we derive three key insights:

\begin{itemize}[leftmargin=*]
    \item \textbf{In the GSU stage}, simple cosine similarity between multimodal embeddings is sufficient for effective retrieval; more complex similarity scoring mechanisms (e.g., attention or joint ID--multimodal fusion) bring negligible gains.
    \item \textbf{In the ESU stage}, explicit multimodal sequence modeling (e.g., through SimTier~\cite{SimTier}) significantly improves performance, and further gains are achieved by fusing multimodal semantics into ID-based attention (via our proposed SA-TA).
    \item \textbf{Representation quality matters more in the ESU than in the GSU}, with multimodal embeddings that capture fine-grained item semantics (e.g., SCL~\cite{SimTier}) yielding the strongest results.
\end{itemize}
Guided by these insights, we propose \textbf{MUSE} (\textbf{MU}ltimodal \textbf{SE}arch-based framework for lifelong user interest modeling)---a simple yet effective paradigm that unifies our findings into a deployable system (see Figure~\ref{fig:framework}).  
Specifically, MUSE employs frozen SCL-based multimodal embeddings~\cite{SimTier} for both stages: 
\textbf{(1) In the GSU stage}, it retrieves the most top-$K$ relevant behaviors via lightweight multimodal cosine similarity;  
\textbf{(2) In the ESU stage}, it fuses ID and multimodal signals for lifelong user interest modeling through a dual-path architecture design---Semantic-Aware Target Attention (SA-TA) enhances ID-based attention with high-quality semantic guidance, while SimTier compresses the multimodal cosine similarity sequence into a histogram to capture semantic relevance.

Beyond the algorithmic contribution, we make two additional practical contributions:  
First, we share our experience with production deployment. MUSE has been deployed in Taobao display advertising system since mid-2025. Integrated into the online serving pipeline, it incurs negligible latency overhead while efficiently processing lifelong user behavior sequences of up to \textbf{100K-length} and delivering substantial top-line metrics gains. 
Second, to foster community research, we open-source \textbf{the first large-scale public dataset featuring lifelong user behavior sequences paired with high-quality multimodal embeddings}, collected from real-world user traffic of Taobao displaying advertising system.

In summary, our contributions are:
\begin{itemize}[leftmargin=*]
    \item We provide the first systematic analysis of multimodal integration in lifelong user interest modeling, revealing distinct design principles for the GSU (simplicity) and ESU (richness + fusion).
    \item We propose MUSE, a practical framework that embodies these principles and achieves state-of-the-art performance in both offline and online evaluations.
    \item We release industrial deployment practices and a large-scale multimodal lifelong behaviors dataset to support future research.
\end{itemize}

%% file: sections/2.Preliminaries.tex
\section{Preliminaries}\label{sec:preliminaries}
This section introduces key concepts and notations used throughout the paper, with a focus on lifelong user behavior sequences modeling for CTR prediction and the multimodal representations employed in our framework.

\paragraph{CTR Prediction with Lifelong Behavior Sequences.}
Click-Through Rate (CTR) prediction estimates the probability that a user will click on a displayed target item, typically formulated as a binary classification task. In industrial-scale recommendation systems, users often have ultra-long historical behavior sequences (e.g., millions of interactions). Given a user $u$ with his behavior sequences $\mathbf{B}_u = [b_1, \dots, b_L]$ ($L$ can be very large), a target item $a$, and other features $o$, the goal is to predict:
\begin{equation}
\label{eq:ctr_prob}
    P(y=1 \mid \mathbf{B}_u, a, o),
\end{equation}
where $y \in \{0,1\}$ indicates a click label. 

Modern architectures (e.g., SIM~\cite{SIM} and TWIN~\cite{TWIN}) address the lifelong sequence challenge via a two-stage framework design:
(1) a \textbf{General Search Unit (GSU)} that retrieves a short, target-relevant subsequence $\mathbf{B}_u^* \subset \mathbf{B}_u$, and  
(2) an \textbf{Exact Search Unit (ESU)} that performs fine-grained modeling on $\mathbf{B}_u^*$ for the final CTR prediction.  
Our work focuses on enhancing both the GSU and ESU stages with rich multimodal information.

\paragraph{Multimodal Item Representations.}
To leverage rich multimodal information of items (e.g., item images and titles), we consider three pre-trained multimodal embedding methods, all of which map an item to a fixed-dimensional vector.

\begin{itemize}[leftmargin=*]
    \item \textbf{OpenCLIP}~\cite{chinese-clip}: An open-source Chinese-adapted CLIP model~\cite{CLIP} pre-trained on 200M image–text pairs via contrastive learning. It provides strong off-the-shelf universal semantic embeddings but lacks user interaction signals.
    
    \item \textbf{I2I (Item to Item)}: A contrastive model trained on item co-occurrence patterns derived from Taobao’s user behavior logs. Positive pairs consist of  frequent item–item transitions, while negatives are sampled via MoCo~\cite{MoCo}. This approach effectively injects collaborative signals into multimodal embeddings.
    
    \item \textbf{SCL (Semantic-aware Contrastive Learning)}~\cite{SimTier}: Constructs positive pairs from user search–purchase behavior (e.g., query image $\leftrightarrow$ purchased item image). Trained with the InfoNCE loss~\cite{InfoNCE}, SCL learns embeddings that capture both semantic signals and behavioral relevance and has proven effective for CTR prediction.
\end{itemize}

All three methods produce \textit{frozen} multimodal embeddings, which are accessed via table lookup during inference. 

%% file: sections/3.Insights.tex
\section{Simple in GSU, Sophisticated in ESU: Principles for Multimodal Lifelong Modeling}
\label{sec:insight}
In this section, we investigate \textbf{how to fully leverage multimodal information to enhance lifelong user interest modeling}, with a systematic focus on integrating multimodal representations into both the GSU and ESU stages. Through extensive experiments on industrial-scale data, we derive the following key insights:

\begin{tcolorbox}[colback=gray!5, colframe=black!30, boxrule=0.5pt, arc=3pt, title={\textbf{Key Insights}}, fonttitle=\bfseries]
\small
\begin{itemize}[leftmargin=*, nosep]
    \item \textbf{GSU:} Multimodal similarity \textit{outperforms} ID-based retrieval, yet \textit{complexity brings little gain}---simple inner-product suffices.
    \item \textbf{ESU:} Multimodal sequence modeling \textit{significantly helps}, and \textit{ID–multimodal fusion} yields further gains.
    \item \textbf{Design Principle:} \textit{Simple GSU + enhanced ESU} is optimal; ESU is far more sensitive to representation quality.
\end{itemize}
\end{tcolorbox}

\subsection{Simple Multimodal Similarity Suffices for Effective GSU Retrieval}
\label{sec:gsu_analysis}
The primary role of the GSU is to efficiently retrieve user behaviors most relevant to the target item from lifelong historical sequences. We analyze two critical design choices: (1) the type of representation  used for similarity computation, and (2) the complexity of the retrieval mechanism.

\begin{table}[!tbp]
    \centering
    \caption{GSU performance with different embeddings. The ESU is the same for all methods for fair comparison.}
    \label{tab:gsu_analysis_1}
    \begin{tabular}{l|cc}
        \toprule
        \textbf{Embedding Type} & \textbf{GAUC} & \textbf{$\Delta$} \\
        \midrule
        ID & 0.6356 & {-} \\
        \midrule
        OpenCLIP & 0.6365 & +0.14\% \\
        I2I & 0.6370 & +0.22\% \\
        SCL & 0.6377 & +0.33\% \\
        \bottomrule
    \end{tabular}
\end{table}

\paragraph{Multimodal-based GSU vs. ID-based GSU}
We compare the ID-based GSU against variants using multimodal representations---namely OpenCLIP, I2I, and SCL—as described in Section~\ref{sec:preliminaries}. Specifically, we compute the cosine similarity between the target item and each historical behavior using the respective embedding types, and retrieve the top-$K$ most similar behaviors as the GSU output.

As shown in Table~\ref{tab:gsu_analysis_1}, \textbf{GSUs based on multimodal representations consistently outperform the ID-based baseline}, demonstrating that semantic signals from multimodal embeddings better capture user–item relevance than discrete IDs.

\begin{table}[htbp]
    \centering
    \caption{Impact of retrieval complexity in the GSU. All variants use SCL embeddings for multimodal similarity.}
    \label{tab:gsu_analysis_2}
    \begin{tabular}{l|cc}
        \toprule
        \textbf{Relevance Metric} & \textbf{GAUC} & \textbf{$\Delta$} \\
        \midrule
        Multimodal cosine & 0.6377 & {-} \\
        \midrule
        Multimodal attention score & 0.6369 & -0.13\% \\
        Multimodal cosine + ID cosine & 0.6379 & +0.03\% \\
        Multimodal cosine + ID attention score & 0.6376 & -0.01\% \\
        \bottomrule
    \end{tabular}
\end{table}

\paragraph{Does increased modeling complexity in the GSU improve performance?}
We explore two enhancement strategies:  
(1) applying a learnable MLP to transform multimodal embeddings before similarity computation (denoted as \textit{multimodal attention score});  
(2) performing joint retrieval by fusing ID-based scores (denoted as \textit{ID Cosine or ID attention score}) and multimodal semantic similarities.

The results are summarized in Table~\ref{tab:gsu_analysis_2}. For (1), we find that \textbf{attention mechanisms---though effective for ID-based features---are not directly transferable to multimodal similarity modeling and lead to noticeably degraded performance}.  
For (2), a naive fusion approach (computing and summing normalized cosine similarities in both ID and multimodal spaces) yields little gain. Even with learnable fusion weights to combine ID-based attention scores and multimodal similarities, the method achieves performance comparable to the simplest multimodal inner-product baseline while incurring higher computational overhead.

These findings lead us to the conclusion that when suitable multimodal representations are available, \textbf{increasing the complexity of the retrieval mechanism does not necessarily yield performance gains}. Instead, a lightweight inner-product similarity suffices for effective GSU retrieval.

\subsection{Multimodal Representations Significantly Enhance the ESU}
\label{sec:esu_analysis}
While prior work has explored multimodal representations to improve the GSU, most approaches still rely solely on ID-based representations in the ESU---typically aggregating historical behaviors via Target Attention---while neglecting the potential of multimodal signals in fine-grained sequence modeling.  
In this work, we investigate two key questions:  
(1) Can multimodal representations improve ESU performance?  
(2) Does increasing modeling complexity in the ESU (e.g., through multimodal–ID fusion) yield additional gains, or is a simple design sufficient---as observed in the GSU?

To address these questions, we analyze two critical design choices:  
(1) replacing the ID-only ESU with a multimodal-only ESU, and 
(2) further fusing multimodal and ID representations in the ESU, analogous to the joint retrieval strategies explored in the GSU.


\begin{table}[htbp]
    \centering
    \caption{Comparison of different ESU designs, evaluating the individual contributions of multimodal modeling (SimTier) and ID-multimodal fusion (SA-TA). All variants use the same GSU and SCL embeddings for fair comparison.}
    \label{tab:esu_analysis_2}
    \begin{tabular}{l|cc}
        \toprule
        \textbf{ESU Type} & \textbf{GAUC} & \textbf{$\Delta$} \\
        \midrule
        Target Attention & 0.6301 & {-} \\
        \midrule
        SimTier & 0.6345 & +0.70\% \\
        Target Attention + SimTier & 0.6361 & +0.95\% \\
        \midrule
        SA-TA & 0.6339 & +0.60\% \\
        SA-TA + SimTier & 0.6377 & +1.21\% \\
        \bottomrule
    \end{tabular}
\end{table}

\paragraph{Multimodal-based ESU vs. ID-based ESU}
For multimodal sequence modeling in the ESU, we adopt SimTier~\cite{SimTier} as a representative architecture. As shown in Table~\ref{tab:esu_analysis_2}, we evaluate the individual and combined effects of SimTier and the standard ID-based Target Attention module.  
The results demonstrate that SimTier delivers significant performance gains---both when used alone and when combined with Target Attention---\textbf{highlighting the substantial benefits of explicitly modeling multimodal semantics in the ESU}.

\paragraph{Fusing multimodal and ID representations yields further gains.}
We explore the fusion of multimodal and ID representations in the ESU, analogous to the joint retrieval strategies examined in the GSU. Specifically, we integrate ID-based attention scores with multimodal semantic similarity within the Target Attention mechanism to produce a unified, semantic-aware attention weight. We refer to this enhanced approach as \textit{Semantic-Aware Target Attention} (SA-TA), with details provided in Section \ref{sec:ESU}.

As shown in Table~\ref{tab:esu_analysis_2}, the proposed SA-TA consistently outperforms the original ID-only Target Attention in the ESU, demonstrating that \textbf{multimodal and ID representations provide complementary signals that can be effectively fused to enhance fine-grained modeling of lifelong user interests}.

\begin{figure}[htbp]
    \centering
    \includegraphics[width=\linewidth]{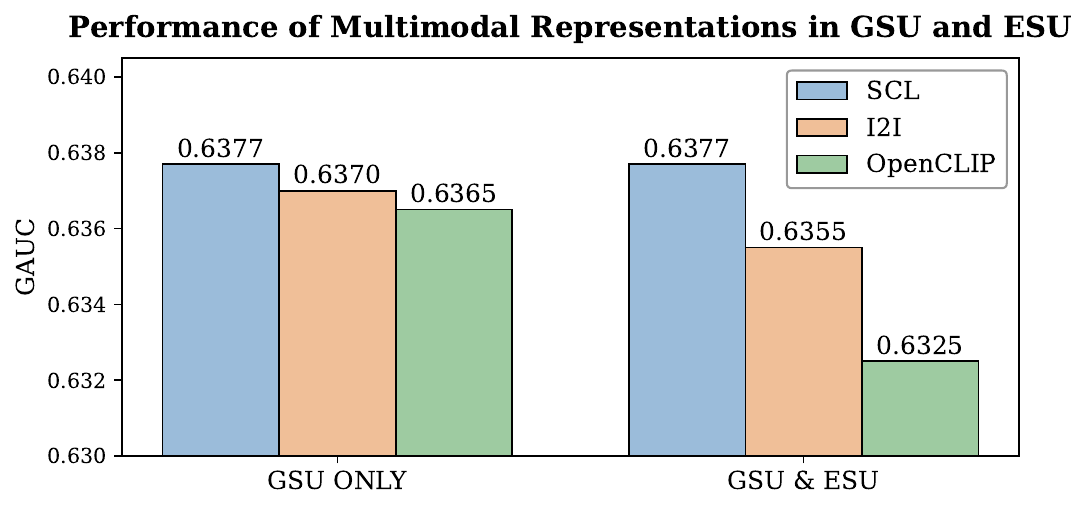}
    \caption{Performance of different multimodal representations. ESU clearly favors fine-grained representations.}
    \vspace{-2ex}
    \label{fig:representation_compare}
\end{figure}

\subsection{Multimodal Representation Quality Matters---Especially in the ESU}
\paragraph{High-quality representations that capture fine-grained item semantics yield better performance.}
As shown in Table~\ref{tab:gsu_analysis_1}, model performance follows the ranking SCL > I2I > OpenCLIP. This is because SCL embeddings capture finer-grained multimodal semantics~\cite{SimTier}, leading to more accurate GSU retrieval and stronger ESU modeling. 
We visualize the differences among these representations  through a GSU case study in Appendix~\ref{sec:case}.

\paragraph{Representation quality has a significantly larger impact on the ESU than on the GSU}
As shown in Figure~\ref{fig:representation_compare}, we quantify this effect by evaluating different multimodal representations under two settings: (1) varying the representation in GSU while keeping the same SCL representation in ESU (GSU ONLY); and (2) jointly replacing the representation in both GSU and ESU (GSU \& ESU). When only the GSU representation is changed, performance remains relatively stable. In contrast, varying the representation in ESU leads to significant performance differences, with fine-grained SCL representaions yielding clear gains. This demonstrates that \textbf{the ESU is far more sensitive to multimodal representation quality than the GSU}, underscoring the importance of semantic modeling in ESU. 

%% file: sections/4.Method.tex
\section{MUSE: A Simple Yet Effective Multimodal Lifelong Modeling Framework}
\label{sec:method}

Guided by the insights from Section~\ref{sec:insight}, we propose \textbf{MUSE} (\textbf{MU}ltimodal \textbf{SE}arch-based framework for lifelong user interest modeling), which adheres to three key principles:  
(1) leveraging high-quality SCL representations,  
(2) employing a \textit{simple, similarity-based GSU}, and  
(3) designing a \textit{multimodal-enhanced ESU} that performs both explicit multimodal sequence modeling and effective ID–multimodal fusion.  
This design strikes an effective balance between efficiency and modeling power. An overview is illustrated in Figure~\ref{fig:framework}.

\subsection{Choice of Multimodal Representations}
As demonstrated in Section~\ref{sec:insight}, representation quality critically impacts performance---especially in the ESU. Among the candidates (OpenCLIP, I2I, SCL), SCL yields the strongest results due to its fine-grained semantic awareness. Therefore, MUSE adopts SCL embeddings for both GSU retrieval and ESU modeling.

\subsection{Multimodal-Based GSU}
Consistent with our finding that \textit{simple inner-product similarity suffices in the GSU}, MUSE uses frozen SCL embeddings to compute semantic relevance via cosine similarity.

Specifically, given a target item $a$ and the lifelong historical behavior sequences of a user $\mathbf{B}_u = [b_1, \dots, b_L]$, we first map each item to its pre-trained multimodal embedding through lookup, resulting in $v_a$ and $\mathbf{V}_u = [v_1, \dots, v_L]$. 

The GSU computes the similarity between the target item and each behavior item:
\begin{equation}
\label{eq:cosine_sim}
    r_i = \langle v_a, v_i \rangle.
\end{equation}
We then select the top-$K$ behaviors with the highest $r_i$ scores to form the subsequence $\mathbf{B}_u^{*}$. This lightweight retrieval aligns with our insight that complex mechanisms bring no gain when high-quality multimodal embeddings are available.

\subsection{Multimodal-Enhanced ESU}\label{sec:ESU}

In contrast to the GSU, our analysis shows that the ESU \textit{requires explicit multimodal sequence modeling and effective fusion of ID and semantic signals} to fully leverage lifelong behavior data. To this end, MUSE’s ESU consists of two complementary components:  
(1) a \textbf{multimodal sequence modeling module} (SimTier) that captures fine-grained semantic relevance patterns across the retrieved behavior sequence, and  
(2) a \textbf{multimodal semantic-aware ID attention module} (SA-TA) that enhances traditional ID-based attention with multimodal semantic guidance.

\paragraph{Explicit Multimodal Sequence Modeling via SimTier.}
Rather than processing raw multimodal embeddings, SimTier~\cite{SimTier} operates on the \textit{semantic similarity sequence} $\mathbf{R} = [r_1, \dots, r_K]$ between the target item and the GSU-retrieved behaviors $\mathbf{B}_u^{*}$. This design aligns with our insight that the ESU benefits from modeling the \textit{structure of semantic relevance} rather than individual embeddings.
Specifically, the similarity range $[-1, 1]$ is uniformly partitioned into $N$ tiers (bins). SimTiermaps similarity sequences $\mathbf{R}$ into an $N$-dimensional histogram $h^{\text{MM}} \in \mathbb{R}^N$, where each entry counts how many similarity scores fall into the corresponding tier. This histogram serves as a compact yet expressive representation of the user’s multimodal interest distribution:
\begin{equation}
\label{eq:lifelong_mm_rep}
 h^{\text{MM}} = \text{Histogram}(\mathbf{R}).
\end{equation}
This approach constitutes our \textit{explicit multimodal sequence modeling}---it captures multi-level semantic patterns (e.g., how many behaviors are highly/weakly relevant) in a single vector.
\begin{figure}[t]
\centering 
\includegraphics[width=.87\columnwidth]{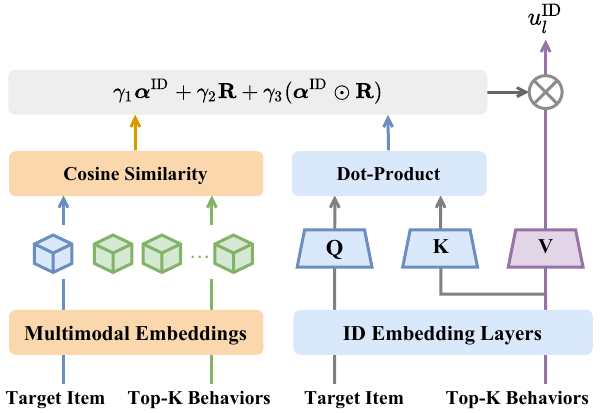} 
\caption{Semantic-Aware Target Attention (SA-TA) augments ID-based attention by incorporating multimodal semantic similarity.} 
\label{fig:SA-TA} 
\end{figure}
\paragraph{Semantic-Aware Fusion in ID-Based Attention (SA-TA)}
While SimTier effectively captures multimodal semantics, the ID path remains essential for capturing collaborative filtering signals. However, as noted in Section~\ref{sec:insight}, ID-only attention suffers from poor generalization on long-tail items within lifelong user behaviors.

To address this, we propose \textit{Semantic-Aware Target Attention} (SA-TA) as shown in Figure~\ref{fig:SA-TA}, which fuses ID-based attention scores with multimodal semantic similarity signals. Given ID embeddings $e_a \in \mathbb{R}^D$ (target) and $\mathbf{E}_u \in \mathbb{R}^{K \times D}$ (behaviors), the standard Target Attention~\cite{DIN} computes by:
\[
\boldsymbol{\alpha}^{\text{ID}} = \frac{(\mathbf{E}_u W^k)(e_a W^q)^\top}{\sqrt{D}}.
\]
SA-TA augments this with the multimodal similarity vector $\mathbf{R}$ (from Eq.~\ref{eq:cosine_sim}) to form fused attention scores:
\begin{equation}
    \boldsymbol{\alpha}^{\text{Fusion}} = \gamma_1 \boldsymbol{\alpha}^{\text{ID}} + \gamma_2 \mathbf{R} + \gamma_3 (\boldsymbol{\alpha}^{\text{ID}} \odot \mathbf{R}),
\end{equation}
where $\gamma_1, \gamma_2, \gamma_3$ are learnable scalars. The final ID-based lifelong user interest representation is obtained by:
\begin{equation}
u_l^{\text{ID}} = \text{Softmax}(\boldsymbol{\alpha}^{\text{Fusion}})^\top (\mathbf{E}_u W^v).
\end{equation}
This design realizes our insight that \textit{ID and multimodal signals are complementary}---SA-TA uses multimodal semantic similarity to enhance ID attention, especially for sparse items.

Finally, the complete lifelong user interest representation is formed by concatenating both paths:
\[
u_l = \left[ h^{MM}, u_l^{\text{ID}} \right],
\]
which is fed into the prediction tower for CTR prediction.

In summary, MUSE’s ESU directly implements the two key findings from Section~\ref{sec:insight}:  
(1) \textit{explicit multimodal sequence modeling} significantly improves over ID-only baselines, and  
(2) \textit{fusing ID and multimodal signals} in attention to achieve further gain.

%% file: sections/5.Deployment.tex
\section{System Deployment}
Industrial-scale recommender systems must return personalized results from corpora of billions of items within tight latency budgets---typically a few hundred milliseconds. 
To meet this constraint, production systems typically employ a multi-stage pipeline: a candidate generation (matching) stage that retrieves a manageable set on the order of $10^3$ items, followed by a compute-heavy ranking stage that applies fine-grained modeling to the candidates.
\begin{figure}[!t]
\centering 
\includegraphics[width=.95\columnwidth]{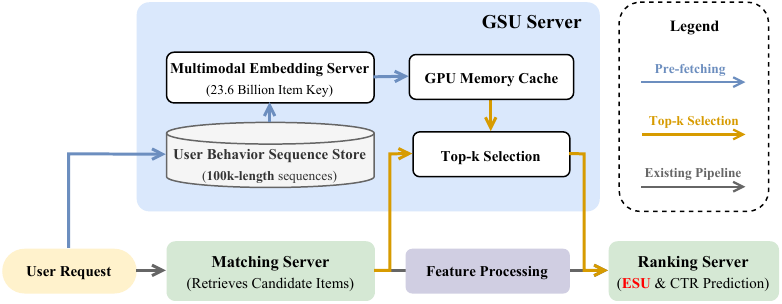} 
\caption{Online deployment of MUSE in Taobao display advertising system. GSU pre-fetches the user behavior sequence and multimodal embeddings asynchronously alongside the matching stage, and the cached outputs are consumed by GSU Top-K selection and ESU modeling during ranking.}
\vspace{-2ex}
\label{fig:production} 
\end{figure} 
In Taobao display advertising system, we deploy MUSE in the ranking stage to effectively model lifelong user behavior sequences with up to 100K interactions. 
The primary performance bottleneck  arises from the substantial network communication required to fetch these long behavior sequences and their embeddings in GSU, which can cause the system to exceed its latency budget. 
Note that this bottleneck is independent of the candidate items, since the user is fixed once the request arrives. 
To eliminate it, we decouple GSU feature/embedding fetching from the critical ranking path, execute it asynchronously in parallel with the matching stage, and cache the embeddings in GPU memory. In the ranking stage, we perform similarity computation, \textbf{Top-$K$ Selection} and  ESU modeling:
\begin{itemize}[leftmargin=*]
    \item \textbf{Pre-fetching of GSU.} In parallel with matching, the GSU server retrieves the user behavior sequence and the multimodal embeddings from remote storage and caches them in GPU memory. Pre-fetching typically completes faster than matching, so its latency is fully hidden.
    \item \textbf{Top-$K$ Selection of GSU}: During ranking, the model uses the cached  embeddings to compute similarities and selects the top-$K$ most relevant historical interactions to feed into the ESU. Note that the similarity computation and the top-$K$ selection can be performed in parallel with other feature-processing operations in the ranking stage, incurring negligible latency.
    \item \textbf{ESU modeling.} Given the selected top-$K$ sub-sequence, the ranking model applies SimTier and SA-TA within the ESU and produces the final CTR prediction.
\end{itemize}

With the aforementioned asynchronous design, MUSE imposes negligible incremental latency in production. Its efficient two-stage design keeps resource consumption within budget.  Since its deployment in mid-2025, MUSE has handled the majority of traffic, delivering stable performance and substantial gains.

%% file: sections/6.Dataset.tex
\section{Dataset Construction}



The MUSE framework targets industrial-scale recommendation scenarios that require modeling lifelong user interests through ultra-long behavior sequences and rich multimodal item content. Such capabilities are essential for real-world systems like Taobao display advertising platform. 

However, existing public recommendation datasets~\cite{KuaiRec,microlens,Tenrec} typically lack either long user behavior sequences or comprehensive multimodal features. To thoroughly evaluate and demonstrate the effectiveness of MUSE, we conduct experiments on two self-constructed datasets derived from Taobao display advertising system, both featuring long user behavior sequences and multimodal item representations. 

The first one is capped by our \textit{production dataset}, fully reflecting real-world traffic and used to validate MUSE’s performance in an industrial setting. The second is a curated \textit{academic dataset}, carefully sampled and anonymized for research purposes. \textbf{It is notable that we will open-source this high-quality dataset to foster community research on multimodal recommendation and lifelong user interest modeling.} 


\begin{table}[t]
    \centering
    \caption{Statistics of industrial production and open-source academic datasets. Here, \textit{Distinct Items} denotes the number of unique items covered by user behaviors and target items. }
    \label{tab:dataset}
    \begin{tabular}{l|cc}
    \toprule
        \textbf{Field} & \textbf{Production} & \textbf{Open-Source}  \\ 
    \midrule
        Samples           & 3.71B      & 99.0M       \\ 
        Users            & 0.19B      & 8.79M       \\ 
        Distinct Items            & 23.6B      & 35.4M      \\ 
    \midrule
        Max Behaviors Length       & 100K  & 1K       \\ 
    \bottomrule
    \end{tabular}
\end{table}

\subsection{Industrial Production Dataset}
The industrial dataset is constructed from impression logs in Taobao display advertising system. For our experiments, we use one week of data: the first six days for training and the final day for testing. Each sample in this dataset comprises hundreds of carefully engineered features. As summarized in Table~\ref{tab:dataset}, the dataset contains 3.71B samples from 0.19B users featuring with 100K historical behavoirs. This dataset enables rigorous offline evaluation of lifelong modeling approaches like MUSE.

\subsection{Open-Source Academic Dataset}
To promote reproducible research, we release a privacy-preserving, open-source academic dataset. It includes a concise set of key features and user behavior sequences of up to several thousand interactions. Due to copyright restrictions on product images and titles, we do not provide raw multimodal content. Instead, we provide \textbf{SCL-based multimodal embeddings}~\cite{SimTier} for all items---high-quality representations that capture both semantic and behavioral relevance, as validated in Section~\ref{sec:insight}. 

The dataset comprises the following components:
\begin{itemize}[leftmargin=*] 
    \item \textbf{User features}: anonymized user ID, age, gender, city, and province.
    \item \textbf{Item features}: item ID, category, item city, and province.
    \item \textbf{Behavior sequences}: A user’s historical sequence consists of up to 1K behaviors, each identified by its item ID. Each item is represented solely by a 128-dimensional SCL embedding. 
    \item \textbf{Label}: a binary label indicating click (1) or non-click (0).
\end{itemize}
As Shown in Table~\ref{tab:dataset}, the open-source dataset contains 107M samples from 8.86M users, covering 275M unique items. The maximum sequence length is capped at 1K to balance research utility and distribution feasibility.

We believe this dataset that combines \textit{long user behavior sequences} with \textit{high-quality multimodal embeddings} will enable systematic exploration of multimodal lifelong interest modeling. The dataset, along with preprocessing scripts and baseline implementations, is now publicly available\footnote{Dataset available at \url{https://huggingface.co/datasets/TaoBao-MM/Taobao-MM}}\footnote{Code available at \url{https://github.com/TaoBao-MM/MUSE}}.

%% file: sections/7.Experiments.tex
\section{Experiments}
\label{sec:experiments}
We conduct comprehensive experiments to evaluate MUSE in the context of CTR prediction. 
\subsection{Experimental Settings}

\subsubsection{Baselines.} We compare MUSE against the following state-of-the-art and widely adopted lifelong interest modeling methods:
\begin{itemize}[leftmargin=*]
    \item \textbf{DIN}~\cite{DIN}: This method introduces a Target Attention mechanism to model user interest from historical behaviors and is the most widely adopted model for short-sequence interest modeling.
    \item \textbf{SIM-Hard}~\cite{SIM}: This method adopts a two-stage approach for long-sequence modeling. In the GSU stage, it retrieves the top-$K$ behaviors that share the same category as the target item. In the ESU stage, it applies DIN to model user fine-grained interest.
    \item \textbf{SIM-Soft}~\cite{SIM}: This variant replaces the category-based retrieval in the GSU stage of SIM-hard with embedding-based similarity search, while still employing DIN in the ESU stage for user fine-grained interest modeling.
    \item \textbf{TWIN}~\cite{TWIN}: This method proposes an efficient Target Attention mechanism applicable to both the GSU and ESU stages, enabling attention-based retrieval already in the GSU stage.
    \item \textbf{MISS}~\cite{MISS}: This approach employs a hybrid retrieval strategy: an ID-based Co-GSU and a multimodal-based MM-GSU. In the ESU stage, it uses ID-based DIN for user interest modeling.
\end{itemize}

\subsubsection{Implementation Details.} For a fair comparison, all methods share the same network architecture, including embedding layers and the upper MLP tower, except for the long-term behavior modeling module. 
All methods consistently use SCL~\cite{SimTier} as the multimodal representation. Each model is trained in one epoch~\cite{ZhangSZJHDZ2022OneEpoch,zong2025recis}.

For DIN, we use the most recent 50 behaviors due to its bottleneck in processing long sequences.
For two-stage models (SIM, TWIN, MISS, MUSE), the GSU retrieves 50 behaviors from the lifelong behavior sequence for ESU. The full behavior sequence length is capped at 5,000 for the production dataset and 1,000 for the open-source dataset. We reproduce each method’s GSU strictly according to the original papers. 
For SIM-soft and SIM-hard, we adopt the same ESU as MUSE (SA-TA and SimTier). Considering that TWIN uses a special variant of Target Attention to improve computational efficiency, we maintain its original design and integrate SimTier into TWIN’s ESU. For MISS, we keep the original ID-only ESU to illustrate the significant effect of multimodal-enhanced ESU.

On the open-source dataset, we train all models using Adam for all parameter groups: embedding layers are trained with a learning rate of 2.0e-3, while DNN parameters use a learning rate of 2.0e-4. The batch size is 8,000.

\subsubsection{Evaluation Metrics.} We adopt Group AUC (GAUC)---a metric demonstrated to align more closely with online metric~\cite{DIN,ShengZZDDLYLZDZ2021STAR,ShengGCYHDJXZ2023JRC}. 




\subsection{Overall Performance}

Table ~\ref{tab:overall_performance} shows the performance of all models on both production and open-source datasets, where MUSE consistently achieves the best results. These comparisons further validate the insights established in Section~\ref{sec:insight}, confirming the effectiveness of our method.

\textbf{First, multimodal-based GSU outperforms ID-based GSU.} When paired with the same ESU, MUSE achieves better performance than both SIM-hard and SIM-soft, which rely on ID-based similarity search.
\textbf{Second, complex GSU mechanisms hardly bring any improvement.} TWIN employs an attention mechanism to model target-behavior relevance but achieves limited performance, partly because attention scores based on ID embedding inner products are unreliable, especially for long-tail items. Recall that in Section~\ref{sec:gsu_analysis}, we also use the fused attention scores of SA-TA in GSU, introducing additional computational overhead without outperforming MUSE’s simpler multimodal semantic similarity search.
\textbf{Finally, multimodal-enhanced ESU brings significant improvement.} Although MISS introduces an additional multimodal GSU stage, it does not utilize multimodal information in the ESU stage, leading to suboptimal results.

\begin{table}[t]
    \centering
    \caption{Overall performance on the production and open-source dataset. Here, to accelerate experimentation,
user behavior sequences in the production dataset are truncated from 100K interactions
to 5K. The \textit{Base} model is set to \textit{DIN}.}
    \label{tab:overall_performance}
    
    \begin{tabularx}{\linewidth}{l|>{\centering\arraybackslash}X>{\centering\arraybackslash}X|>{\centering\arraybackslash}X>{\centering\arraybackslash}X}
        \toprule
        \multirow{2}{*}[-0.5ex]{\textbf{Method}} 
            & \multicolumn{2}{c|}{\textbf{Production-5k}} 
            & \multicolumn{2}{c}{\textbf{Open-Source-1k}} \\
        \cmidrule(lr){2-3} \cmidrule(lr){4-5}
              & \textbf{GAUC} & \textbf{$\Delta$} & \textbf{GAUC} & \textbf{$\Delta$} \\ 
        \midrule
        Base            & 0.6271        & {-}           & 0.6082         & {-}             \\
        \midrule
        SIM-Hard        & 0.6351        & +1.27\%       & 0.6139          & +0.94\%             \\
        SIM-Soft        & 0.6356        & +1.35\%       & 0.6135          & +0.87\%             \\
        TWIN            & 0.6304        & +0.53\%       & 0.6098          & +0.26\%             \\
        MISS            & 0.6308        & +0.59\%       & 0.6097          & +0.25\%                 \\
        \midrule
        MUSE            & \textbf{0.6377}& \textbf{+1.69\%} & \textbf{0.6154}        & \textbf{+1.18\%}            \\
        \bottomrule
    \end{tabularx}
\end{table}

\subsection{Impact on Behavior Sequence Length}
\begin{figure}[!t]
    \centering
    \includegraphics[width=\linewidth]{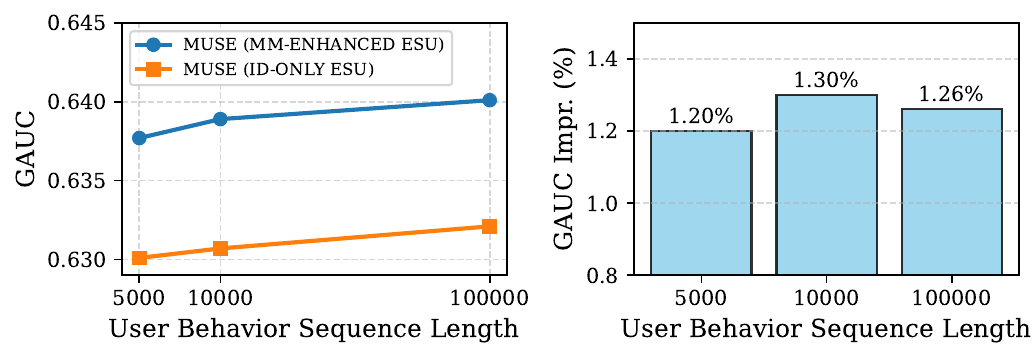}
    \caption{Performance when GSU Takes Different Behavior Sequence Lengths. Left: GAUC values. Right: Relative GAUC improvement of MUSE using MM-Enhanced ESU compared to ID-Only ESU at different sequence lengths.}
    \vspace{-2ex}
    \label{fig:seq_length}
\end{figure}

We evaluate the effectiveness of MUSE under varying behavior sequence lengths. As shown in Figure~\ref{fig:seq_length}, we observe that (1) MUSE performance improves significantly as sequence length increases, \textbf{extending from 5K to 100K behaviors yields a +0.38\% gain in GAUC}; and (2) across all lengths, the multimodal-enhanced ESU \textbf{consistently outperforms the ID-only ESU by a large margin}. This aligns with our insight  that the ESU benefits from high-quality multimodal representations.

\subsection{Performance on Different Users and Items}
\begin{figure}[t]
    \centering
    \begin{minipage}{\linewidth}
        \centering
        \includegraphics[width=\linewidth]{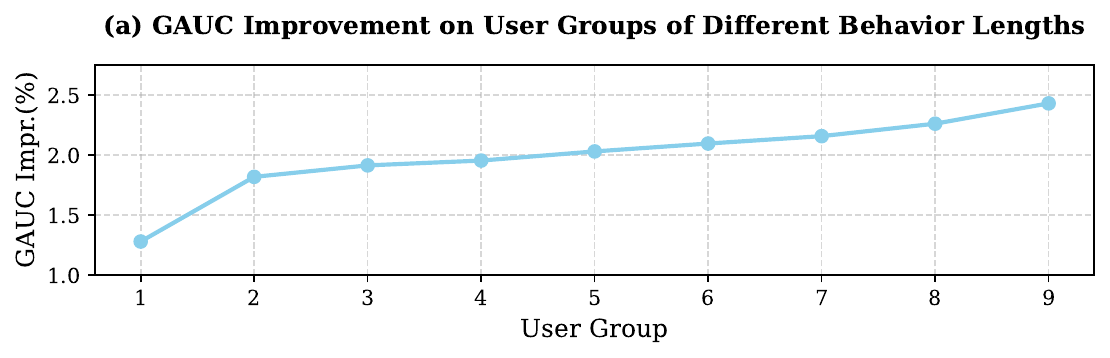}
    \end{minipage}
    \begin{minipage}{\linewidth}
        \centering
        \includegraphics[width=\linewidth]{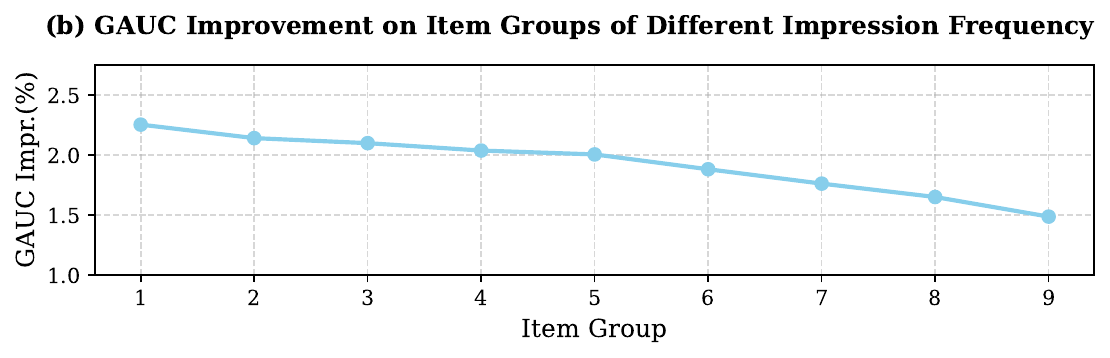}
    \end{minipage}
    \vspace{-2ex}
    \caption{Relative GAUC improvement across user groups divided by behavior length, and item groups divided by impression frequency.}
    \vspace{-2ex}
    \label{fig:user_item_group_gauc}
\end{figure}

To evaluate MUSE across user activity levels, we partition users into nine equal-sized groups by behavior sequence length. As shown in Figure~\ref{fig:user_item_group_gauc} (a), relative GAUC improvement increases monotonically from Group 1 (shortest) to Group 9 (longest), confirming that \textbf{MUSE benefits most from rich behavioral context}.

In a similar manner, we divide items into nine equally-sized groups, from Group 1 (least popular) to Group 9 (most popular), and compute the relative GAUC improvement per group. As shown in Figure~\ref{fig:user_item_group_gauc} (b), MUSE achieves higher gains on niche items, demonstrating \textbf{strong generalization with multimodal information}.





\subsection{Online A/B Test Result}
\begin{table}[htbp]
    \centering
    \caption{The online performance lift of the proposed MUSE framework compared with the production baseline.}
    \label{tab:performance_lift}
    \begin{tabular}{l|ccc}
    \toprule
        \textbf{Metric} & \textbf{CTR} & \textbf{RPM} & \textbf{ROI} \\ 
    \midrule
        Lift            & +12.6\%      & +5.1\%       & +11.4\%      \\ \bottomrule
    \end{tabular}

\end{table}
We implemented MUSE (with sequence length of 100K) in Taobao display advertising system in mid-2025 and evaluated its online performance through a long-term A/B test. The production baseline was based on SIM with behavior length of 5K. As shown in Table~\ref{tab:performance_lift}, the deployment of MUSE with the 100K-length version yielded substantial performance improvements, \textbf{achieving gains of +12.6\% in CTR (Click-Through Rate), +5.1\% in RPM (Revenue Per Mille) and +11.4\% in ROI (Return On Investment)}. 

%% file: sections/8.Related.tex
\section{Related Work}
\label{sec:related}

Our work intersects two active research directions: (1) lifelong user interest modeling, and (2) multimodal recommendation.

\subsection{Lifelong User Interest Modeling}
Modeling ultra-long user behavior sequences is crucial for capturing evolving interests. Early approaches such as MIMN~\cite{MIMN} employ memory networks to maintain user interest states over time.  
A major breakthrough comes from the two-stage framework introduced by SIM~\cite{SIM} and UBR4CTR~\cite{UBR4CTR}, which decouples retrieval and modeling: a \textbf{General Search Unit (GSU)} efficiently retrieves $K$ most relevant behaviors from the full history, and an \textbf{Exact Search Unit (ESU)} performs fine-grained sequence modeling on this subsequence for final prediction.

Subsequent works refine the GSU design: ETA~\cite{ETA} uses locality-sensitive hashing with Hamming distance for fast retrieval; SDIM~\cite{SDIM} groups items by hash signatures; TWIN~\cite{TWIN} extends Target Attention~\cite{DIN} to the GSU, selecting behaviors with the highest attention scores.  
However, since attention scores are optimized for \textit{behavior aggregation}, not \textit{behavior selection}, making them suboptimal for retrieval~\cite{DARE}. More fundamentally, \textbf{all these methods rely solely on ID-based features}, inheriting well-known limitations: poor generalization for long-tail items and lack of semantic expressiveness.  
These shortcomings motivate recent efforts---including our work---to integrate multimodal signals into lifelong modeling.


\subsection{Multimodal Recommendation}
Multimodal information (e.g., images, text) has shown great promise in enhancing recommendation accuracy~\cite{DVBPR,BM3,MMGCN,DeepCTR,wheretogo}. 

AlignRec~\cite{AlignRec} aligns fixed CLIP features with ID embeddings under collaborative filtering supervision. Sheng et al.~\cite{SimTier} propose Semantic-aware Contrastive Learning (SCL), constructing behavior-grounded positive pairs (e.g., search query $\leftrightarrow$ purchased item) during pre-training, and then model user interests via SimTier and MAKE using these fixed representations.  
QARM~\cite{QARM} aligns multimodal features with item-item relationships during pre-training and enables end-to-end optimization through quantized codes~\cite{RQVAE,Tiger}.

Most relevant to our work, MISS~\cite{MISS} pioneers multimodal integration in lifelong behavior modeling by introducing a multimodal GSU alongside the ID-based one. However, \textbf{MISS discards multimodal signals in the ESU}, restricting semantic enrichment to the GSU retrieval and leaving the modeling stage vulnerable to ID embedding limitations.  

In contrast, we propose a \textbf{systematic integration of multimodal information across both GSU and ESU}. Our analysis reveals distinct design principles for each stage---simplicity suffices in the GSU, while the ESU benefits from explicit multimodal sequence modeling and ID–semantic fusion---leading to a simple yet highly effective framework, MUSE.

%% file: sections/9.Conclusion.tex
\section{Conclusion}
\label{sec:conclusion}

In this work, we systematically investigate how to effectively integrate multimodal representations into lifelong user interest modeling. Through extensive experiments, we uncover distinct design principles for the two-stage architecture: the General Search Unit (GSU) benefits from simplicity---high-quality multimodal embeddings with inner-product similarity suffice---while the Exact Search Unit (ESU) requires explicit multimodal sequence modeling and effective fusion of ID and semantic signals. Guided by these insights, we propose MUSE, a simple yet effective framework that achieves state-of-the-art performance in both offline and online evaluations. MUSE has been deployed in Taobao display advertising system since mid-2025, delivering significant top-line business metrics with negligible latency overhead. Furthermore, we release industrial deployment practices and the first large-scale open-source dataset featuring ultra-long behavior sequences paired with SCL-based multimodal embeddings to support future research. We hope our analysis, framework, and resources will inspire more effective and efficient approaches to multimodal-enhanced lifelong interest modeling in both academia and industry.

%% file: sections/10.Acknowledgements.tex
\section*{Acknowledgements}\label{sec:ACKNOWLEDGEMENTS}
We sincerely thank Jing Huang, Jinjing Li, Jiawen Liao, Zhenyuan Lai, Wenchao Wang, Chaochao Zhao, Yunlong Xu, Zhengxiong Zhou, Huimin Yi, Xingyu Wen, Dun Yang, Yan Zhang, Jinzhe Shan, Gaoming Zhou, Xiang Gao, Rui Du, Xiaorui Zhang, Qifeng Li, Jiamang Wang, Peng Sun, and other colleagues for their invaluable engineering support on MUSE.

%% file: sections/11.Appendix.tex
\section{Appendix}

\subsection{GSU Case Study}\label{sec:case}
We present three cases in Figure~\ref{fig:gsu_case} to illustrate the distinct search results of different GSU types across diverse target items (i.e., children's sun-protection clothes, hairpin and mobile Wi-Fi):
\begin{itemize}[leftmargin=*]
    \item \textbf{MUSE with SCL}: The multimodal GSU with SCL embeddings precisely captures visual details that distinguish individual items, exhibiting fine-grained semantic matching capabilities. It accurately retrieves historical items that that closely resemble the target item in style, shape, color, texture, etc.
    \item \textbf{MUSE with OpenCLIP}: In contrast, GSU based on OpenCLIP embeddings searches at a coarser granularity, attending to overall image content. It often retrieves items based on irrelevant background or contextual similarity rather than the item itself. For instance, in case 1 the child model, in case 2 the trademark, and in case 3 the text “3000G” are overemphasized during retrieval.
    \item \textbf{SIM-hard}: The category-based GSU correctly identifies items within the same category as the target item, but overlooks fine-grained attributes such as color and style, which are often the aspects users care about most.
    \item \textbf{SIM-soft}: The ID-based GSU prioritizes items that frequently co-occur with the target item in users' interaction histories, resulting in high popularity bias and low semantic relevance. For example, children’s toys in case 1, women’s products in case 2, and electronic devices in case 3 are all retrieved primarily due to co-occurrence rather than semantic similarity to the target item.
\end{itemize}